\documentclass{ws-ijmpa}

\begin{document}

\def\nocropmarks{\vskip5pt\phantom{cropmarks}}

\let\trimmarks\nocropmarks    

\markboth{V.~Lensky {\it et al.}}
{Advances in the field theoretic understanding of $NN\to NN\pi$}

%
\catchline{}{}{}{}{}
%

\title{ADVANCES IN THE FIELD-THEORETIC UNDERSTANDING OF PION PRODUCTION
  IN NUCLEON-NUCLEON COLLISIONS}

\author{VADIM LENSKY\footnote{also Institute for Theoretical and Experimental Physics,
  Bol'shaya Cheremushkinskaya 25, Moscow, 115275 Russia.}, JOHANN
  HAIDENBAUER, CHRISTOPH HANHART
}
\address{Institut f\"ur Kernphysik, Forschungszentrum J\"ulich GmbH,
D-52425 J\"ulich, Germany
}
\author{VADIM BARU, ALEXANDER KUDRYAVTSEV}
\address{Institute for Theoretical and Experimental Physics,
  Bol'shaya Cheremushkinskaya 25,
Moscow, 117259 Russia}
\author{ULF-G. MEI\ss NER\footnote{also Institut f\"ur Kernphysik, Forschungszentrum J\"ulich GmbH,
D-52425 J\"ulich, Germany}}
\address{Helmholtz-Institut f\"ur Strahlen- und Kernphysik (Theorie),
  Universit\"at Bonn, Nu\ss allee 14-16, D-53115 Bonn, Germany}
\maketitle

\begin{abstract}
We study the production amplitude for the reaction $NN\to NN\pi$ up to next-to-leading
order in chiral perturbation theory.
We show that the irreducible chiral loops at this order exactly cancel those
terms that arise from the off-shell parts of the $\pi N$ rescattering vertex. This cancellation is
required for formal consistency of the whole scheme.
The net effect of the inclusion of all next-to-leading order loops is to enhance the leading
rescattering amplitude by a factor of 4/3 compared to phenomenological
studies, bringing its contribution to the cross section for
$pp\to d\pi^+$ close to the experimental value.

\end{abstract}

\ccode{PACS numbers: 21.30.Fe, 12.39.Fe, 25.10.+s, 25.40.Ve}

\section*{}
Understanding the dynamics of pion production in nucleon-nucleon collisions near
threshold is of significant importance. It is the first hadronic
inelasticity of $NN$ scattering at intermediate energies, thus we can
only understand $NN$ scattering if we understand $NN\to NN\pi$.
It is also a necessary step to an understanding of isospin
violation in few-nucleon processes\cite{vKMiNi00,Gaa04},
which provides a test for chiral perturbation theory (ChPT), and it is
needed for a calculation of absorptive corrections to $\pi d$ scattering\cite{VL06_1}.
When accurate data for the total cross-section close to threshold appeared in 1990\cite{IUCF90}, existing models\cite{KR,Mi90} failed
to describe the data by a factor of five to ten for the channel
$pp\to pp\pi^0$ and a factor two for the channels $pp\to pn\pi^+$ and
$pp\to d\pi^+$.
To cure this discrepancy, many mechanisms were proposed -- for a recent review
see Ref.~\refcite{CH04}.

In the reaction $NN\to NN\pi$ the momentum transfer $|\vec p|$ is large compared to the pion mass already at threshold:
$|\vec p_{\mathrm{thr}}|=\sqrt{m_\pi M}$ with $m_\pi\, (M)$ --- the pion (nucleon) mass. This new scale has to be accounted for,
which leads to modifications of Weinberg's counting scheme\cite{We90,We92}. 
The expansion parameter in this case is\cite{Co96,dR00,CH00,CH02} 
\begin{equation}
\chi=\frac{|\vec p_{\mathrm{thr}}|}{M}=\sqrt{\frac{m_\pi}{M}}.
\end{equation}

\begin{figure}[b]
\centerline{\epsfig{file=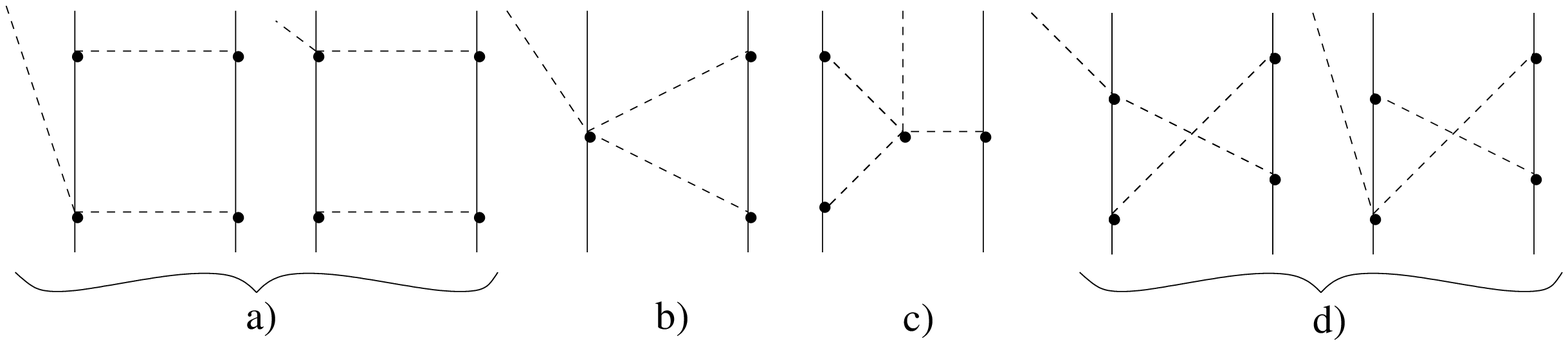,width=10.7cm,angle=0}}
\vspace*{8pt}
\caption{Leading loop diagrams for $NN\to NN\pi$. Solid lines are
  nucleons, dashed lines are pions.\label{diagram}}
\end{figure}\noindent
In Ref.~\refcite{CH02} the large
momentum scale was shown to promote some loops to lower order compared to
Weinberg's original counting. The leading loops
are shown in Fig.\ref{diagram}; the loops (b)--(d) enter at
next-to-leading order (NLO). The diagrams (a) in Fig.\ref{diagram} are reducible according to the
common rules -- the one-pion-exchange is regarded as part of the wave
function. Therefore they were not included into the transition
operator.
The findings of Ref.~\refcite{CH02} were:
\begin{itemlist}
\item For the channel $pp\to pp\pi^0$ the sum of diagrams (b)--(d) of Fig.
  \ref{diagram} canceled:
\begin{equation}
A_{pp\to pp\pi^0}^{1b+1c+1d}= \frac{g_A^3}{256f_\pi^5}\left(-2+3-1\right)\, 
|\vec p|=0;
\end{equation}
\item For the channel $pp\to d\pi^+$ the same sum gave a finite answer:
\begin{equation}
A_{pp\to d\pi^+}^{1b+1c+1d} = \frac{g_A^3}{256f_\pi^5} \left(-2+3+0\right)\,
|\vec p|=\frac{g_A^3|\vec p|}{256f_\pi^5}.
\end{equation}
\end{itemlist}\noindent
The latter amplitude grows linearly with increasing final
 $NN$--relative momentum, which leads to a large
  sensitivity to the final $NN$ wave function, once the convolution of those
  with the transition operators is evaluated. However, such a
 sensitivity is not allowed in a consistent field theory. This problem was stated
 in Ref.~\refcite{Gaa05}; to cure this, a new counterterm at leading order
 was proposed. However, such a structure would violate chiral symmetry.
We show here how to resolve this problem. At the same time we shed
new light on the concept of reducibility in pion reactions on few-nucleon
systems. For the details see Ref.~\refcite{VL05_2}.
Our central finding is that the diagrams (a) in Fig.\ref{diagram} in fact contain a genuine irreducible piece due to the
energy dependence of the leading order (LO) $\bar NN\pi\pi$
vertex. This irreducible part should be considered along with the
diagrams (b)--(d) of Fig.~\ref{diagram}. Without loss of generality we
work in threshold kinematics, {\it i.e.}, we assume all final
particles at rest and the initial relative momentum to be $\vec p$.
We get for the full expression for the first diagram (a) of
Fig.\ref{diagram} up to higher orders:
\begin{eqnarray}\label{Eqq}
A^{1a1}_{pp\to d\pi^+}\!\!\!\!&\quad =\quad &\!\!\!\!i\frac{3g_A^3}{32f_\pi^5}
\int \frac{d^4l}{(2\pi)^4}
\frac{[l_0+m_\pi-(2\vec p+\vec l)\cdot \vec l/(2M)]}
{(l_0-\frac{m_\pi}{2}-\frac{(\vec l+\vec
  p)^2}{2M}+i\epsilon)(-l_0+\frac{m_\pi}{2}-\frac{(\vec l+\vec
  p)^2}{2M}+i\epsilon)}\quad
 \\ \nonumber
& & \qquad\qquad \qquad \qquad \qquad \qquad \qquad \times
\frac{(\vec l\cdot(\vec l+\vec p))}{
(l^2-m_\pi^2)((l+p)^2-m_\pi^2)} \ .
\end{eqnarray}
Let us rewrite the expression for the $\bar N N \pi\pi$ vertex in
the numerator in Eq.~\ref{Eqq} (note: since
$\vec p\,^2/M,\, \vec l\,^2/M\sim m_\pi$, the recoil term contributes to the vertex at LO as well):
\begin{equation}
\left[l_0+m_\pi-\frac{(2\vec p+\vec l)\cdot \vec l}{2M}\right]=
\left[\left(l_0-\frac{m_\pi}{2}-\frac{(\vec p+\vec l)^2}{2M}\right)
+2m_\pi\right],\label{Eqm}
\end{equation}
where we used threshold kinematics. The first term in the
r.h.s. of Eq.~\ref{Eqm} cancels the first nucleon propagator in
Eq.~\ref{Eqq}. The corresponding piece of the diagram is
irreducible and enters at NLO.
We get for its contribution up to
higher orders:
\begin{equation}
A^{1a1 (\mbox{irr})}_{pp\to d\pi^+}=-\frac{3}{4}\frac{g_A^3|\vec p|}{256f_\pi^5} \ ,
\label{irr1a1}
\end{equation}
where the label $(\mbox{irr})$ indicates that this is only the irreducible piece of
the diagram.
Analogous considerations apply to the second diagram of diagrams (a) of
Fig.~\ref{diagram}. Their contribution to the reaction
$pp\to pp\pi^0$ is zero due to the isovector character of the leading
$\bar NN\pi\pi$ vertex. Thus, one gets for the sum of all the NLO contributions:
\begin{eqnarray} 
\nonumber
A_{pp\to d\pi^+}^{1a1(\mbox{irr})+1a2(\mbox{irr})+1b+1c+1d} &=& 
\frac{g_A^3}{256f_\pi^5}\left(-\frac{3}{4}-\frac{1}{4}-2+3+0\right)\,
|\vec p|=0 \ ,\\
A_{pp\to pp\pi^0}^{1a1(\mbox{irr})+1a2(\mbox{irr})+1b+1c+1d} &=& 
\frac{g_A^3}{256f_\pi^5}\left(\phantom{-}0\,\,+0\,\,-2+3-1\right)\,
|\vec p|=0 \ .
\label{cancel}
\end{eqnarray}
Thus, in both channels that contribute at the production threshold the sum of all
irreducible NLO loops cancels. And no counterterm is necessary at this
order, at variance with the claims of Ref.~\refcite{Gaa05}.

\begin{figure}[t]
\centerline{\epsfig{file=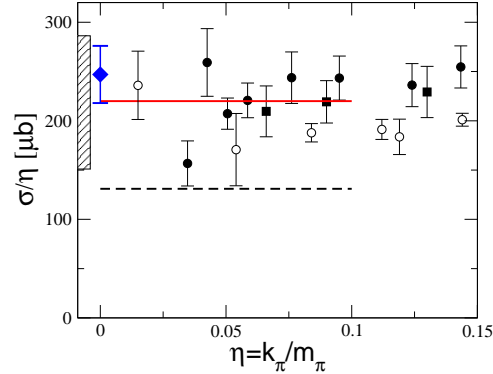,height=7.7cm,angle=270}}
\vspace*{8pt}
\caption{Comparison of our results to experimental data for $pp\to d\pi^+$.
The dashed curve shows the LO results. The solid line shows the
results at NLO. The hatched bar shows the theoretical uncertainty for
the NLO result. Data are from Refs.~\protect\refcite{dpidata1}
 (open circles), \protect\refcite{dpidata2} (filled circles) and \protect\refcite{dpidata3} (filled
  squares). The diamond shows result obtained from the width of pionic
  deuterium\protect\cite{dgotta}.\label{f1}
}
\end{figure}

The remaining
pieces in the expressions for $A_{pp\to d\pi^+}^{1a}$ exactly agree to the convolution
of the LO rescattering contribution with the $NN$ wave function, however,
with the $\bar N N\pi \pi$ vertex put on-shell, {\it i.e.} instead of the commonly
used\cite{KR} $3/2\ m_\pi$ in the vertex we have to use the
value $2\,m_\pi$ -- cf. Eq.~\ref{Eqm}. This enhances the dominating isovector $\pi
N$-rescattering amplitude by a factor of $4/3$, which leads to a good
description of the experimental data for $pp\to d\pi^+$ --- see Fig.~\ref{f1} for the comparison.

To summarize, some pion production diagrams
that seem reducible contain in fact irreducible pieces as a result
of the energy dependence of the LO $\bar N N\pi \pi$
vertex. For the reaction $pp\to d\pi^+$
the net effect of the inclusion of all NLO loops is to enhance the LO
rescattering amplitude by a factor of $4/3$, bringing its contribution
to the cross section close to the experimental value.
The NLO contributions that would lead to a large
sensitivity to the off-shell parameters cancel, which is a necessity
for the formalism to be consistent.

\end{document}